\documentclass[prd,showpacs,12pt]{revtex4}
\usepackage{amsmath,amssymb,revsymb,graphicx,dcolumn}

\newcommand{\beq}{\begin{equation}}
\newcommand{\eeq}{\end{equation}}
\newcommand{\beqa}{\begin{eqnarray}}
\newcommand{\eeqa}{\end{eqnarray}}
\newcommand{\bsubeqs}{\begin{subequations}}
\newcommand{\esubeqs}{\end{subequations}}

\input{tcilatex}
\begin{document}

\title{Comment on "Creation of spin 1/2 particles by an electric field in de
Sitter space" }
\author{\textbf{S. Haouat}}
\email{s.haouat@gmail.com}
\affiliation{\textit{LPTh, Department of Physics, University of Jijel, BP 98, Ouled
Aissa, Jijel 18000, Algeria.}}
\author{\textbf{R. Chekireb}}
\affiliation{\textit{LPTh, Department of Physics, University of Jijel, BP 98, Ouled
Aissa, Jijel 18000, Algeria.}}

\begin{abstract}
The main objective of this comment is to correct the density number of
created fermions by an electric field in the (1+1) dimensional de-Sitter
space-time communicated by V. M. Villalba in Phys. Rev. D \textbf{52,} 3742
(1995). It is shown how the positive and negative energy solutions are
connected to one another by the charge conjugation. Some concluding remarks
are made.
\end{abstract}

\maketitle

Several years ago, V. I. Villalba studied the problem of spin half particle
creation by an electric field in (1+1) dimensional de-Sitter space-time. In
the first part of his paper Villalba has derived the pair creation rate in
the absence of the electric field. Then he considered in a second part the
effect of the electric field on the particle creation. He has considered a
spin $\frac{1}{2}$\ fermion of mass $m$ and charge $e$ moving in dS$_{2}$
space-time described by the line element 
\begin{equation}
ds^{2}=a^{2}\left( \eta \right) \left( -d\eta ^{2}+dx^{2}\right) ,
\end{equation}%
where $\eta $ is the conformal time $\eta =\frac{-1}{H}e^{-Ht}$ and $a\left(
\eta \right) =\frac{-1}{H\eta }$. For the electric field he has chosen the
gauge $A_{\mu }=$( $0$, $A_{1}\left( \eta \right) $), with 
\begin{equation}
A_{1}=-\frac{E_{0}}{H}e^{Ht}=\frac{E_{0}}{H^{2}\eta }.
\end{equation}%
To accomplish his study Villalba has determined the positive and negative
energy solutions for the corresponding Dirac equation and used these
solutions to compute the rate of particles created via the Bogoliubov
transformation connecting the "in" with the "out" states. As a result,~he
has concluded that the ratio between Bogoliubov coefficients is given by%
\begin{equation}
\left\vert \frac{\alpha }{\beta }\right\vert ^{2}=\frac{\sinh \frac{\pi }{H}%
\left( \mathcal{M}-\frac{eE_{0}}{H}\right) }{\sinh \frac{\pi }{H}\left( 
\mathcal{M}+\frac{eE_{0}}{H}\right) }\exp \left[ \frac{2\pi }{H}\left( 
\mathcal{M}-\frac{eE_{0}}{H}\right) \right] ,  \label{1}
\end{equation}%
where%
\begin{equation}
\mathcal{M}=\sqrt{m^{2}+\frac{e^{2}E_{0}^{2}}{H^{2}}}.  \label{2}
\end{equation}%
Reading carefully that paper, one can see that Villalba used an erroneous
formula between Whittaker functions (see equation (42) in \cite{V}).
Consequently, equation (\ref{1}) is incorrect. In addition, he has not shown
how the positive and negative solutions are connected to one another by the
charge conjugation. This symmetry is more important in the quantization of
the spinor field and the derivation of the particle creation rate.
Furthermore, the author claimed that for the Dirac case, we do not need to
impose any restriction on the relation between $m$ and $H$ in order to
obtain well defined \textquotedblleft out\textquotedblright\ states.
However, this should be investigated rigorously since the particle creation
is well-defined only in the adiabatic regimes. From physical point of view,
this is the major drawback of \cite{V} because the fact that light or
super-heavy particles created in abundant will affect significantly the
cosmic evolution.

Before making any comment let us first use the solutions obtained by
Villalba for the Dirac equation to write the Dirac spinors corresponding to
the "in" states in the form

\begin{equation}
\Psi _{in}^{+}\left( \eta \right) =\mathcal{N}_{in}\left( 
\begin{array}{c}
W_{\lambda ,\mu }\left( 2ik_{x}\eta \right) -i\frac{m}{H}W_{\lambda -1,\mu
}\left( 2ik_{x}\eta \right) \\ 
-iW_{\lambda ,\mu }\left( 2ik_{x}\eta \right) +\frac{m}{H}W_{\lambda -1,\mu
}\left( 2ik_{x}\eta \right)%
\end{array}%
\right)  \label{3}
\end{equation}%
and%
\begin{equation}
\Psi _{in}^{-}\left( \eta \right) =\mathcal{N}_{in}^{\ast }\left( 
\begin{array}{c}
\frac{m}{H}W_{-\lambda ,\mu }\left( -2ik_{x}\eta \right) +iW_{-\lambda
+1,\mu }\left( -2ik_{x}\eta \right) \\ 
-W_{-\lambda +1,\mu }\left( -2ik_{x}\eta \right) -i\frac{m}{H}W_{-\lambda
,\mu }\left( -2ik_{x}\eta \right)%
\end{array}%
\right)  \label{4}
\end{equation}%
$\allowbreak $where $\mathcal{N}_{in}$ is a normalization constant that is
unimportant vis-a-vis the mechanism of particle creation and 
\begin{align}
\mu & =i\frac{\mathcal{M}}{H},  \label{5} \\
\ \lambda & =\frac{1}{2}+\frac{ieE_{0}}{H^{2}}.  \label{6}
\end{align}

\bigskip For the "out" states, we have%
\begin{equation}
\Psi _{out}^{+}\left( \eta \right) =\mathcal{N}_{out}\left( 
\begin{array}{c}
\sqrt{\mathcal{M}+\frac{eE_{0}}{H}}M_{\lambda ,\mu }\left( 2ik_{x}\eta
\right) +\sqrt{\mathcal{M}-\frac{eE_{0}}{H}}M_{\lambda -1,\mu }\left(
2ik_{x}\eta \right) \\ 
-i\sqrt{\mathcal{M}+\frac{eE_{0}}{H}}M_{\lambda ,\mu }\left( 2ik_{x}\eta
\right) +i\sqrt{\mathcal{M}-\frac{eE_{0}}{H}}M_{\lambda -1,\mu }\left(
2ik_{x}\eta \right)%
\end{array}%
\right)  \label{7}
\end{equation}%
and%
\begin{equation}
\Psi _{out}^{-}\left( \eta \right) =\mathcal{N}_{out}^{\ast }\left( 
\begin{array}{c}
\sqrt{\mathcal{M}-\frac{eE_{0}}{H}}M_{-\lambda ,-\mu }\left( -2ik_{x}\eta
\right) -\sqrt{\mathcal{M}+\frac{eE_{0}}{H}}M_{-\lambda +1,-\mu }\left(
-2ik_{x}\eta \right) \\ 
-i\sqrt{\mathcal{M}-\frac{eE_{0}}{H}}M_{-\lambda ,-\mu }\left( -2ik_{x}\eta
\right) -i\sqrt{\mathcal{M}+\frac{eE_{0}}{H}}M_{-\lambda +1,-\mu }\left(
-2ik_{x}\eta \right)%
\end{array}%
\right) ,  \label{8}
\end{equation}%
where $\mathcal{N}_{out}$ is also a normalization constant. Here we note
that those states are connected to one another by the charge conjugation
transformation defined by 
\begin{equation}
\Psi \rightarrow \Psi ^{c}=i\sigma _{2}\Psi ^{\ast }.  \label{9}
\end{equation}%
In addition, positive and negative energy solutions satisfy the
orthogonality condition%
\begin{equation}
\bar{\Psi}^{+}\Psi ^{-}=\bar{\Psi}^{-}\Psi ^{+}=0.  \label{10}
\end{equation}%
Let us now use the Bogoliubov transformation connecting the "in" with the
"out" states to determine the density of created particles. The relation
between those states can be obtained by the use of the relation between
Whittaker functions \cite{Grad} 
\begin{equation}
M_{\lambda ,\mu }\left( \rho \right) =\Gamma \left( 2\mu +1\right) e^{i\pi
\lambda }\left[ \frac{W_{-\lambda ,\mu }\left( -\rho \right) }{\Gamma \left(
\mu -\lambda +\frac{1}{2}\right) }+\frac{W_{\lambda ,\mu }\left( \rho
\right) }{\Gamma \left( \mu +\lambda +\frac{1}{2}\right) }\exp \left[ -i\pi
\left( \mu +\frac{1}{2}\right) \right] \right]  \label{11}
\end{equation}%
with $-\frac{3\pi }{2}<\arg \rho <\frac{\pi }{2}$ and $2\mu \neq -1,-2,\cdot
\cdot \cdot $. The spinor $\Psi _{out}^{+}\left( \eta \right) $ can be then
expressed in terms of $\Psi _{in}^{+}\left( \eta \right) $ and $\Psi
_{in}^{-}\left( \eta \right) $ as follows%
\begin{equation}
\Psi _{out}^{+}\left( \eta \right) =\alpha ~\Psi _{in}^{+}\left( \eta
\right) +\beta ~\Psi _{in}^{-}\left( \eta \right)  \label{12}
\end{equation}%
where the Bogoliubov coefficients $\alpha ~$and $\beta $ are given by 
\begin{equation}
\frac{\alpha }{\beta }=\frac{\mathcal{N}_{in}}{\mathcal{N}_{in}^{\ast }}%
\frac{\Gamma \left( \frac{1}{2}+\mu -\lambda \right) }{\Gamma \left( \frac{1%
}{2}+\mu +\lambda \right) }\sqrt{\mathcal{M}^{2}-\frac{e^{2}E_{0}^{2}}{H^{2}}%
}e^{-i\pi \left( \mu +\frac{1}{2}\right) }  \label{13}
\end{equation}%
and%
\begin{equation}
\left\vert \alpha \right\vert ^{2}+\left\vert \beta \right\vert ^{2}=1.
\label{14}
\end{equation}%
By the use of the following property of gamma function \cite{Grad} 
\begin{equation}
\left\vert \Gamma (ix)\right\vert ^{2}=\frac{\pi }{x\sinh \pi x}  \label{15}
\end{equation}%
we find%
\begin{equation}
\left\vert \frac{\alpha }{\beta }\right\vert ^{2}=\frac{\sinh \pi \left( 
\frac{\mathcal{M}}{H}+\frac{eE_{0}}{H^{2}}\right) }{\sinh \pi \left( \frac{%
\mathcal{M}}{H}-\frac{eE_{0}}{H^{2}}\right) }e^{2\pi \frac{\mathcal{M}}{H}}.
\label{16}
\end{equation}%
Notice that equation (\ref{16}) differs from equation (\ref{1}). As is
mentioned above, this is because the author of \cite{V} used an erroneous
formula between Whittaker functions.

For the density number of created particles we have $n\left( k\right)
=\left\vert \beta \right\vert ^{2}$. Taking into account that the Bogoliubov
coefficients satisfy the condition (\ref{14}), it is easy to show that%
\begin{equation}
n\left( k_{x}\right) =\frac{\sinh \pi \left( \frac{\mathcal{M}}{H}-\frac{%
eE_{0}}{H^{2}}\right) e^{-2\pi \frac{\mathcal{M}}{H}}}{\sinh \pi \left( 
\frac{\mathcal{M}}{H}+\frac{eE_{0}}{H^{2}}\right) +\sinh \pi \left( \frac{%
\mathcal{M}}{H}-\frac{eE_{0}}{H^{2}}\right) e^{-2\pi \frac{\mathcal{M}}{H}}}.
\label{17}
\end{equation}%
We note that this result is obtained for a positive wave vector ($k_{x}>0$)
and $-\frac{3\pi }{2}<\arg \left( 2ik_{x}\eta \right) <\frac{\pi }{2}$. For
the the case when $k_{x}<0$, the quantity $n\left( k_{x}\right) $ can be
obtained form (\ref{17}) by changing the sign of $e$.

Let us notice also that the auxiliary functions $\Theta _{1}$ and $\Theta
_{2}$ introduced in \cite{V} satisfy the system of equations 
\begin{align}
\left[ \eta \frac{d}{d\eta }-\frac{1}{2}+i\left( k_{x}\eta -\frac{eE_{0}}{%
H^{2}}\right) \right] \Theta _{1}& =\frac{m}{H}\Theta _{2}  \label{18} \\
\left[ \eta \frac{d}{d\eta }-\frac{1}{2}-i\left( k_{x}\eta -\frac{eE_{0}}{%
H^{2}}\right) \right] \Theta _{2}& =-\frac{m}{H}\Theta _{1},  \label{19}
\end{align}%
where the mixing term vanishes in the limit $m\rightarrow 0$. This means
that for massless particles, the positive and negative energy solutions
never intercept each other and there is no production of massless particles
even if an electric field is present.

Furthermore, since equation (36) in \cite{V} is of the form $\Theta
_{1,2}^{\prime \prime }\left( \eta \right) +\omega _{1,2}^{2}(\eta )\Theta
_{1,2}\left( \eta \right) =0$, with $\omega _{1,2}^{2}(\eta )=\left( \frac{%
\mathcal{M}^{2}}{H^{2}}+\frac{1}{4}\right) \frac{1}{\eta ^{2}}-2k_{x}\left( 
\frac{eE_{0}}{H^{2}}\mp \frac{i}{2}\right) \frac{1}{\eta }+k_{x}^{2}$, we
can see that the adiabatic condition 
\begin{equation}
\lim\limits_{\eta \rightarrow 0}\left\vert \frac{\dot{\omega}_{1,2}(\eta )}{%
\omega _{1,2}^{2}(\eta )}\right\vert \approx \frac{1}{\left( \frac{\mathcal{M%
}^{2}}{H^{2}}+\frac{1}{4}\right) ^{\frac{1}{2}}}<<1  \label{20}
\end{equation}%
implies that $\mathcal{M}>>H$. Taking into account this condition, the
density of created particles can be approximated by 
\begin{equation}
n\left( k_{x}\right) =\exp \left[ -\frac{2\pi }{H}\left( \mathcal{M}+\func{%
sign}\left( k_{x}\right) \frac{eE_{0}}{H}\right) \right] ,  \label{21}
\end{equation}%
which is similar to the density of scalar particles (see Eq. (27) in \cite{G}%
). It follows from (\ref{21}) that $n\left( k_{x}\right) $ is more
significant when $k_{x}<0$. Therefore the constant electric field produces
predominantly particles with $k_{x}<0$. In other wards, in the presence of a
constant electric field, particles prefer to be created with a specific sign
of the canonical momentum $k_{x}$. This depends on the orientation of the
electric field and the sign of the particle charge. The antiparticles are in
the main created with the opposite sign for $k_{x}$.

In conclusion, we have succeeded to derive the exact number density of
created fermions by an electric field in (1+1) dimensional de-Sitter
space-time. We have found that equation (\ref{16}) in the present paper
differs from equation (\ref{1}). This is due to a mistake in equation (42)
in \cite{V} which leads to incorrect results. Let us notice that a constant
electric field amplifies the creation of fermions with negative wave number $%
k_{x}$ and minimizes it in the opposite direction. The inverse is true for
antiparticles. In addition we have shown that the creation of light
particles is possible in the presence of the electric field and thermal
spectrum of particle creation will be obtained only when $\mathcal{M}$
satisfy the adiabatic condition $\frac{\mathcal{M}}{H}>>1$. Another
important result is that the creation of massless particles with conformal
coupling is impossible even if an electric field is present.

\appendix

\section{Useful relations}

The spinors given in this comment are obtained from Villalba's solutions by
the use of the following relations \cite{Vilenkin}

\begin{equation*}
\left[ \rho \frac{\partial }{\partial \rho }+\frac{1}{2}\rho -\lambda \right]
W_{\lambda ,\mu }\left( \rho \right) =\left[ \left( \lambda -\frac{1}{2}%
\right) ^{2}-\mu ^{2}\right] W_{\lambda -1,\mu }\left( \rho \right) ,
\end{equation*}%
\begin{equation*}
\left[ \rho \frac{\partial }{\partial \rho }-\frac{1}{2}\rho +\lambda \right]
W_{\lambda ,\mu }\left( \rho \right) =-W_{\lambda +1,\mu }\left( \rho
\right) ,
\end{equation*}%
\begin{equation*}
\left[ \rho \frac{\partial }{\partial \rho }+\lambda -\frac{1}{2}\rho \right]
M_{\lambda ,\mu }\left( \rho \right) =\left( \mu +\lambda +\frac{1}{2}%
\right) M_{\lambda +1,\mu }\left( \rho \right) ,
\end{equation*}%
and%
\begin{equation}
\left[ \rho \frac{\partial }{\partial \rho }-\lambda +\frac{1}{2}\rho \right]
M_{\lambda ,\mu }\left( \rho \right) =\left( \mu -\lambda +\frac{1}{2}%
\right) M_{\lambda -1,\mu }\left( \rho \right) .
\end{equation}

\end{document}